\documentclass[journal]{IEEEtran}

\usepackage[OT1]{fontenc} 
\usepackage[numbers,sort&compress]{natbib}
\usepackage[cmex10]{amsmath}
\usepackage{amssymb}
\usepackage{bm}
\usepackage{braket}
\usepackage{graphicx}
\usepackage{color}
\usepackage{subfigure}
\usepackage{tabularx}
\usepackage{arydshln}
\usepackage{mathtools}
\usepackage{multirow}
\usepackage{algorithm,algorithmic}
\usepackage{url}
\usepackage{listings}
\usepackage{comment}
\usepackage{CJKutf8}

\def\vec{\mathrm{vec}}

\def\bs{\boldsymbol}

\begin{document}

\title{\LARGE Penalty-Free Two-Step Optimization of Higher-Order Ising Problems for Two-Dimensional Line-Controlled RIS}

\author{
Tasuku~Okamoto,~\IEEEmembership{Graduate Student Member,~IEEE},  Naoki~Ishikawa,~\IEEEmembership{Senior~Member,~IEEE},
Daisuke~Kitayama, Yuto~Hama,~\IEEEmembership{Member,~IEEE},
Kensuke~Inaba,~\IEEEmembership{Member,~IEEE},
Toshimori~Honjo,~\IEEEmembership{Member,~IEEE},
Hiroki~Takesue,~\IEEEmembership{Member,~IEEE},
Hiroyuki~Takahashi,~\IEEEmembership{Member,~IEEE}.
\thanks{T.~Okamoto, N.~Ishikawa, and Y.~Hama are with the Faculty of Engineering, Yokohama National University, Kanagawa 240-8501, Japan (e-mail: iskw@ieee.org).
D.~Kitayama and H.~Takahashi are with Device Technology Laboratories, NTT Inc.; K.~Inaba, T.~Honjo, and H.~Takesue are with Basic Research Laboratories, NTT Inc., Atsugi, Kanagawa 243-0198, Japan.
}}

\markboth{\today}
{Shell \MakeLowercase{\textit{et al.}}: Bare Demo of IEEEtran.cls for Journals}

\maketitle

\begin{abstract}
Reconfigurable intelligent surfaces (RISs) are often assumed to allow continuous phase control over all elements, leading to hardware cost that scales with the number of elements. Treating the phase of each element as a discrete variable is essential for improving cost effectiveness toward ubiquitous RIS deployment. However, the resulting discrete optimization problem is inherently difficult to solve. To address this challenge, this letter proposes a two-dimensional line-control method to reduce the degrees of freedom of the phase variables. The formulation yields a fourth-order objective function and is not directly compatible with physical optimizers such as coherent Ising machines and quantum annealers, which are designed for quadratic interactions. Conventional methods for reducing the order of the objective function with additional auxiliary variables increase the number of variables and require additional penalty parameters, limiting scalability. We therefore propose a two-step optimization method that transforms the fourth-order objective into two successive quadratic optimization problems. For a RIS with 5,476 elements, the required number of discrete variables is reduced from 11,100 to 5,476. Experiments using a real coherent Ising machine demonstrated that the proposed approach solved the discrete-phase optimization problem with 5,476 elements, while limiting the beamforming-gain loss to 2~dB compared with the full continuous-control case.
\end{abstract}


\IEEEpeerreviewmaketitle

\section{Introduction}
\IEEEPARstart{R}{econfigurable} intelligent surfaces (RISs) have attracted significant attention as a promising solution to address the propagation challenges of millimeter-wave and THz bands in next-generation wireless communication systems.
A RIS is composed of a large number of unit elements that control the scattering of electromagnetic waves, thereby enhancing the received signal power at desired locations. 
Additionally, since RISs do not require amplifiers, they offer substantial hardware cost reductions compared to traditional base stations, digital relays, and analog repeaters~\cite{wu2020smart}.
Their compact form factor and flexible deployment capabilities enable the integration into various structures. In particular, RISs composed of transparent substrates are well suited for installation on windows~\cite{kitayama2021transparent}, as they maintain visibility and natural lighting. Moreover, large-scale RISs can be integrated into urban infrastructures such as billboards and outdoor digital signage, enabling new application scenarios. 

A critical design challenge in RIS implementation is reducing the hardware costs associated with phase control~\cite{wu2020smart}. Increasing the number of RIS elements improves beamforming gain, but requires more switching elements, resulting in higher hardware complexity and cost.
Therefore, it is essential to develop control strategies that reduce the hardware complexity without compromising beamforming gain~\cite{kitayama2026transmissive}. 

From a theoretical standpoint, continuous phase control can achieve higher beamforming gain, and efficient optimization methods for continuous phase control, such as the complex circle manifold method, have been proposed~\cite{elmossallamy2021ris}. However, in practical hardware implementations, phase control is typically restricted to discrete levels, such as two-level and four-level control~\cite{wu2020smart,hama2026discrete}, due to limitation in hardware costs and complexity of hardware control.Such discrete phase-control schemes are formulated as binary combinatorial optimization problems that are generally NP-hard. Consequently, achieving high beamforming gain is difficult. To address this challenge, heuristic methods such as genetic algorithms and machine learning have been investigated to obtain approximate solutions~\cite{liu2021reconfigurable}.

Over the past decade, the development of physical machines for solving discrete optimization problems has been actively pursued. Representative machines are the quantum annealer (QA) \cite{amin2015searching} and the coherent Ising machine (CIM) \cite{honjo2021100000spin}. Both machines have been studied for RIS phase-shift optimization \cite{hama2026discrete,zheng2026constrained}. Notably, the CIM provides a highly configurable Ising solver, in which arbitrary quadratic Ising interaction networks can be embedded without graph-conversion overhead. It has been shown to provide high-quality approximate solutions with superior speed and accuracy for large-scale problems compared with QA and classical simulated annealing \cite{hamerly2019experimental}. With a quantum algorithm, a Grover-based approach offers a theoretical quadratic speedup \cite{ishikawa2026quantumaccelerated}, yet it depends on fault-tolerant quantum computers, which remain infeasible. In this context, the CIM represents a practically feasible engineering approach for near-term industrial applications.

Against this background, we propose a two-dimensional line-control method for discrete RIS phase shifts, in which each phase shift is determined by the product of the corresponding row and column control variables. This structure reduces the number of control variables, thereby decreasing hardware cost and facilitating wider deployment. When the proposed line-control method is mapped onto an optimization problem, the resulting objective function includes fourth-order terms in the row and column control variables. This unconstrained fourth-order objective is not directly handled by the framework in \cite{zheng2026constrained}, and thus requires a dedicated quadratization method. We therefore propose a two-step optimization method that quadratizes the objective function. The contributions of this letter are as follows.

\begin{enumerate}
    \item \textit{We propose an efficient phase control method for RISs based on two-dimensional line-control.}
    The proposed method achieves sufficient beamforming performance even under limited phase state. It reduces driver circuit costs and simplifies the control process, thereby contributing to the development of more practical and scalable RIS systems.

    \item \textit{We propose a two-step optimization method for solving fourth-order Ising problems.}
    Compared to conventional quadratization methods, our method reduces the number of optimization variables and removes the need for hyperparameter tuning in the transformation process, thus simplifying implementation and improving solution accuracy.
\end{enumerate}

Unlike the full-element RIS optimization focusing on LoS-dominant scenarios in \cite{hama2026discrete}, this work proposes a two-dimensional line-control architecture that significantly reduces the number of control variables.

\section{System Model}
\label{sec:sys}

\begin{figure}[tb]
\centering
\includegraphics[width=0.9\linewidth]{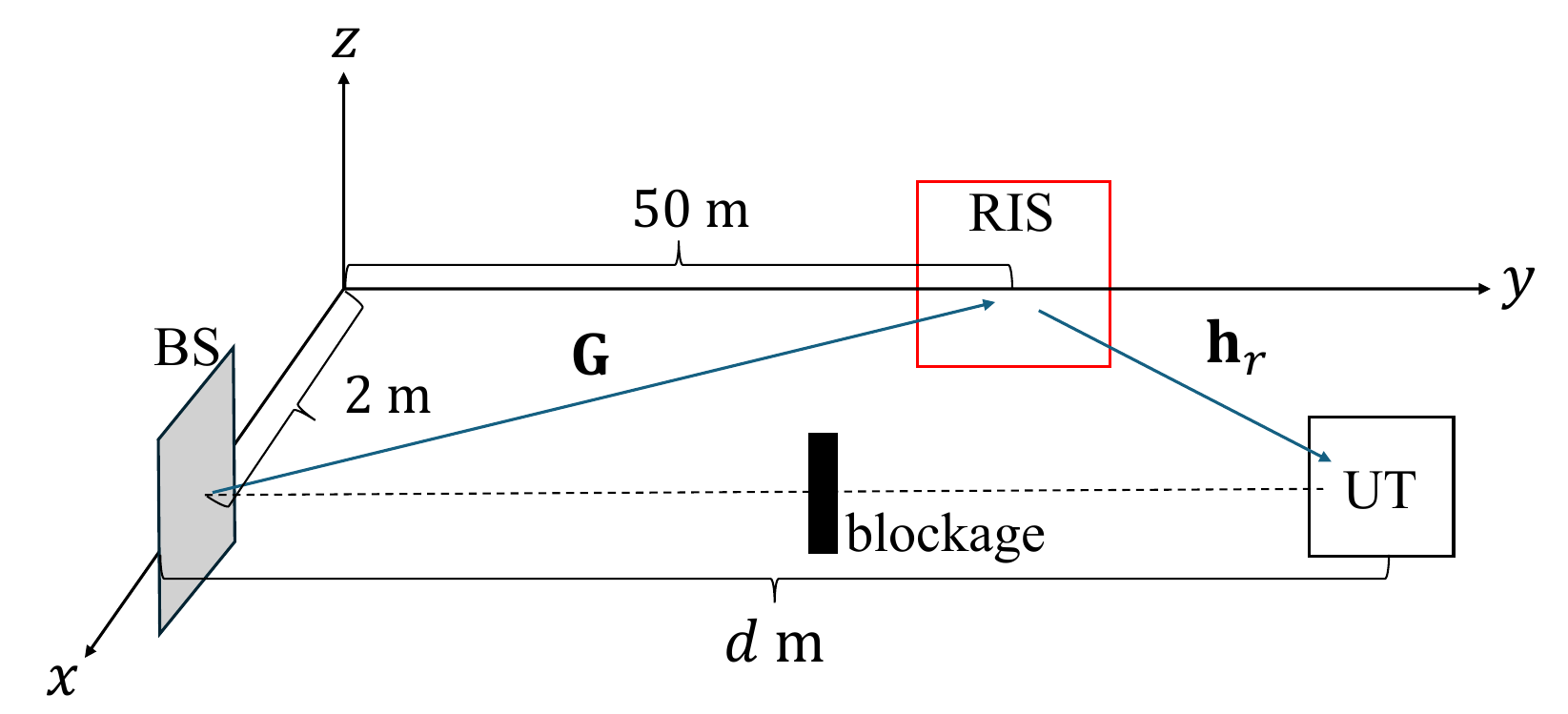}
\caption{The considered RIS-assisted system model with a blocked direct path, where $\mathbf{G}$ and $\mathbf{h_r}$ denote the BS-RIS and RIS-UT channels, respectively.}
\label{fig:line_system} 
\end{figure}

We consider a downlink multiple-input single-output communication system assisted by a RIS with $N$ elements, as illustrated in Fig.~\ref{fig:line_system}.  
In this model, a base station (BS) equipped with $M$ antennas transmits signals to a user terminal (UT) with a single antenna under a non-line-of-sight condition.
Since the relative positions of the BS and the RIS are fixed, the channel coefficients between them are assumed to be quasi-static. Furthermore, we assume a codebook-based operation for the RIS phase shifts, obviating the need for real-time optimization.
The propagation channel from the BS to the RIS is denoted by $\mathbf{G} \in \mathbb{C}^{N\times M}$, and the channel from the RIS to the UT is denoted by $\mathbf{h}_r \in \mathbb{C}^{1\times N}$.  
Using both channel matrices, the received signal $r$ at the UT can be expressed as
\begin{align}
r=\mathbf{h}_r\boldsymbol{\Phi}\mathbf{G}\mathbf{w}s+v,\label{eq:received_symbol}
\end{align}
where $\mathbf{w} \in \mathbb{C}^{M\times1}$ is a beamforming vector at the BS, $s\in \mathbb{C}$ is a transmitted data symbol, and $v\in \mathbb{C}$ is additive white Gaussian noise. The transmitted symbol is assumed to be normalized such that $\mathbb{E}[|s|^2]=1$.
In addition, $\boldsymbol{\Phi}$ is a diagonal matrix representing the phase shifts of the RIS elements, given by
\begin{align} 
\boldsymbol{\Phi}=\mathrm{diag}(\boldsymbol{\phi})=\mathrm{diag}(e^{j\theta_0},e^{j\theta_1},\ldots,e^{j\theta_{N-1}}) \in \mathbb{C}^{N\times N},
\end{align}
where $\theta_i \in [0,2\pi)$ denotes the phase shift of the $i$th RIS element, and thus $\phi_i=e^{j\theta_i}$.
Here, perfect channel state information (CSI) is assumed.
From \eqref{eq:received_symbol}, the received power $P_r$ at the UT can be written as
\begin{align}
P_r =|\mathbf{h}_r\boldsymbol{\Phi}\mathbf{G}\mathbf{w}|^2.\label{eq:power_mimo}
\end{align}
In this letter, the RIS phase shifts $\boldsymbol{\Phi}$ are optimized to maximize the received power at a designated location. 
The optimization problem is formulated as
\begin{align}
    \begin{split}
        \max_{\boldsymbol{\phi}}& \quad P_r,\\
        \textrm{s.t.}& \quad |\phi_i|^2 =1 \quad \forall i.
    \end{split}\label{eq:objective_con}
\end{align}
This problem can be solved as a continuous optimization problem on a complex circle manifold \cite{elmossallamy2021ris}.
When the UT has a single antenna, the BS beamforming vector is optimally determined by maximum ratio transmission as $\mathbf{w} = (\mathbf{h}_r\boldsymbol{\Phi}\mathbf{G})^\mathrm{H}$, independent of the RIS phase distribution \cite{wu2019beamforming}.  
It follows that the objective function can be equivalently expressed as
\begin{align}
(\mathbf{h}_r\boldsymbol{\Phi}\mathbf{G})(\mathbf{h}_r\boldsymbol{\Phi}\mathbf{G})^\mathrm{H}.\label{eq:objective_1}
\end{align}
Here, we consider discrete phase shifts with $L$ levels, $\theta_i\in \{0,2\pi/L,\ldots2\pi(L-1)/L\}$. For $L=2$, i.e., $\phi\in\{-1,+1\}$, the optimization can be simplified by considering only the real part. The problem is then rewritten as 
\begin{align}
    \begin{split}
        \max_{\bs\phi}& \quad \boldsymbol{\phi}^\mathrm{H}\mathbf{J}\boldsymbol{\phi},\\
        \textrm{s.t.}& \quad \phi_i \in\{-1,+1\} \quad \forall i,\label{eq:objective_ising}
    \end{split}
\end{align}
where $\mathbf{J}\in \mathbb{R}^{N\times N}$ is defined as
\begin{align}
\mathbf{J}=\Re\{\mathrm{diag}(\mathbf{h}_r)\mathbf{GG}^\mathrm{H}\mathrm{diag} (\mathbf{h}_r)^\mathrm{H}\}.
\end{align}
Since $\phi_i^2=1$, the objective function can be rewritten as
\begin{align}
\sum_{i<j}J_{i,j}\phi_i\phi_j.
\end{align}
 Then, maximizing the received power of the phase-limited RIS-assisted system is equivalent to solving the Ising problem, which is NP-hard in general, in contrast to the continuous optimization problem of \eqref{eq:objective_con}.

For an extension to $L=4$, the number of spin variables is doubled.  
In this case, the phase shift vector $\boldsymbol{\phi}$ can be expressed as
\begin{align} 
\boldsymbol{\phi}=\frac{1}{\sqrt{2}}(\boldsymbol{\sigma}_{\mathrm{Re}}+j\boldsymbol{\sigma}_{\mathrm{Im}})\in\left\{e^{j\frac{\pi}{4}},e^{j\frac{3\pi}{4}},e^{j\frac{5\pi}{4}},e^{j\frac{7\pi}{4}}\right\}^{N\times 1}. 
\end{align}
where $\boldsymbol{\sigma}_{\mathrm{Re}}, \boldsymbol{\sigma}_{\mathrm{Im}}\in\{+1,-1\}^{N\times1}$.
This formulation enables the realization of a four-level discrete phase shift configuration.

\section{Conventional Techniques}
\label{sec:conv}
This section reviews the CIM, a state-of-the-art physical system for solving discrete optimization problems. It also introduces a standard quadratization method for transforming higher-order Ising problems into quadratic unconstrained binary optimization (QUBO) problems or equivalent Ising problems. 

\subsection{Coherent Ising Machine}
\begin{figure}[tb]
    \centering
    \includegraphics[width=0.87\linewidth]{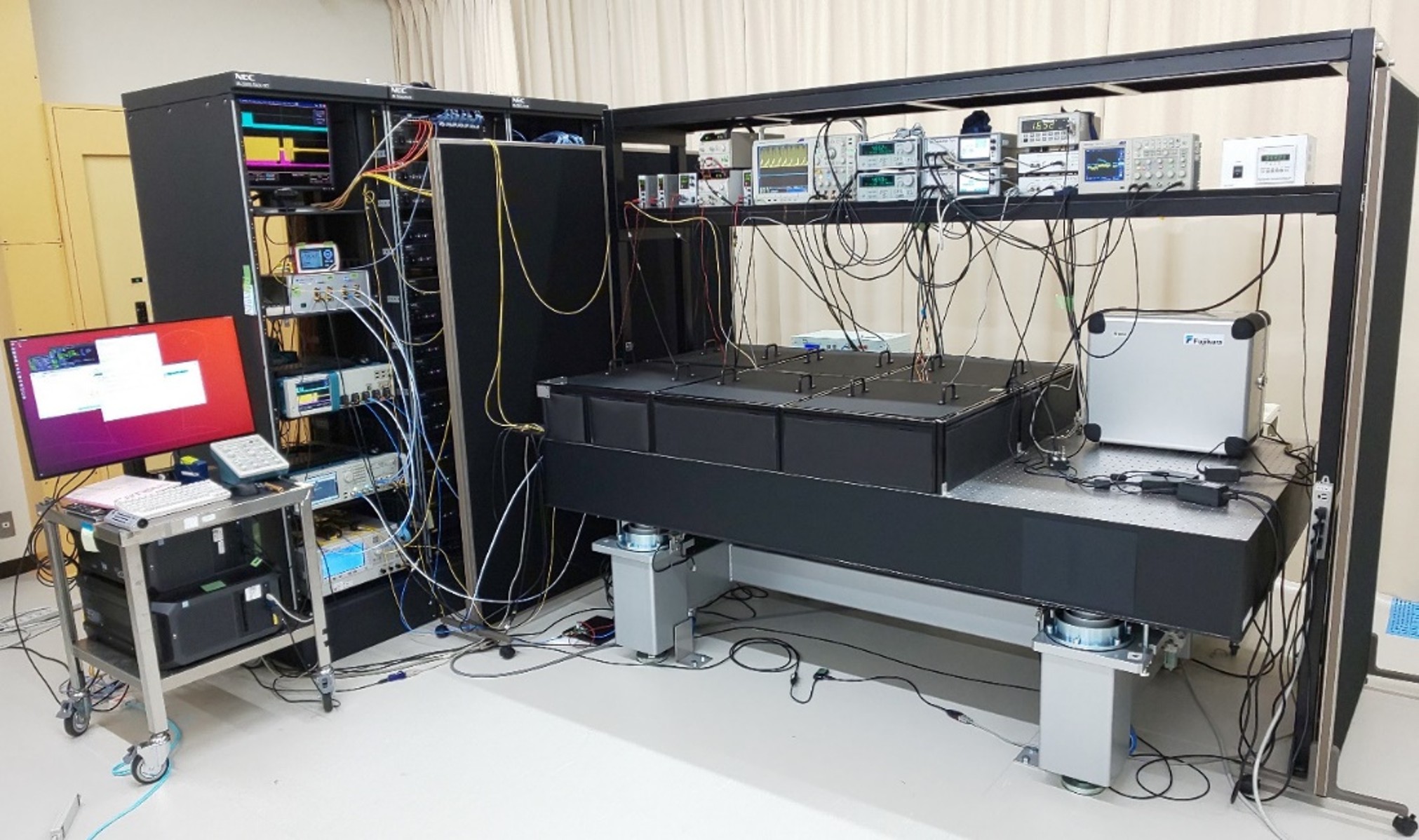}
    \caption{Photograph of the CIM used in our experiments.}
    \label{fig:CIM}
\end{figure}

We use the time-multiplexed degenerate optical parametric
oscillator (DOPO)-based CIM originally developed in~\cite{honjo2021100000spin}. 
In the CIM, each DOPO pulse circulating in a 5-km fiber ring represents an Ising spin through its discretized phase state ($0$ or $\pi$), corresponding to $\sigma_i \in \{-1,+1\}$. 
These binary phase states emerge through phase bifurcation above the oscillation threshold, enabling a direct physical representation of Ising spins. 
Fig.~\ref{fig:CIM} shows the system used in this work.
The CIM solves an Ising problem formulated as
\begin{equation}
H = \sum_i h_i \sigma_i + \sum_{i<j} J_{i,j} \sigma_i \sigma_j, 
\qquad \sigma_i \in \{-1,+1\}.
\label{eq:ising}
\end{equation}

The measurement-feedback architecture in~\cite{honjo2021100000spin} enables programmable pairwise couplings. Then, large-scale, densely connected quadratic instances of the Ising model can be implemented directly in hardware without overhead.
The CIM operates at room temperature and provides approximate solutions through the real-time physical evolution of the optical network. 
Since the CIM deal with quadratic Ising models, higher-order objective functions must be reformulated into quadratic form prior to operation.

\subsection{Quadratization of Higher-Order Ising Problems \cite{mandal2020compressed}}
\label{subsec: quadratrization}
Physical computing machines such as QA and CIM are designed to solve
quadratic problems such as Ising and QUBO formulations, whereas many practical formulations
contain higher-order interactions.
A higher-order Ising objective $H_{\mathrm{orig}.}(\boldsymbol{\sigma})$ 
must be converted into a quadratic counterpart $H_{\mathrm{orig}.}'$ by introducing auxiliary spins 
$\boldsymbol{w}$ that represent intermediate spin products. 
The resulting Hamiltonian is
\begin{align}
H_{\mathrm{tot}.} 
= H_{\mathrm{orig}.}'(\boldsymbol{\sigma},\boldsymbol{w}) 
+ \alpha H_{\mathrm{pen}.}(\boldsymbol{\sigma},\boldsymbol{w}),
\end{align}
where $\alpha H_{\mathrm{pen}.}$ with $\alpha > 0$ enforces consistency between auxiliary spins 
and the corresponding spin products.
In general, this transformation increases the number of spins 
and pairwise couplings. The solution quality strongly depends on the choice of $\alpha$, and the resulting overhead becomes particularly problematic for large and dense problems.

\section{Proposed Line-Control-Based Optimization}
\label{sec:prop}
\begin{figure}[tb]
    \centering
    \includegraphics[width=0.7\linewidth]{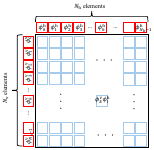}
    \caption{Proposed two-dimensional line-control of a RIS that significantly reduces hardware complexity.}
    \label{fig:ris_line}
\end{figure}
We propose a two-dimensional line-control method for the RISs that enables efficient hardware implementation. For a RIS with $N_{\mathrm{v}}$ rows and $N_{\mathrm{h}}$ columns, the number of control variables is reduced from $N = N_\mathrm{v}N_\mathrm{h}$ to $N_{\mathrm{v}}+N_{\mathrm{h}}$. As illustrated in Fig.~\ref{fig:ris_line}, two sets of phase variables $\boldsymbol{\phi}_\mathrm{v}=(\phi^{\mathrm{v}}_0,\phi^{\mathrm{v}}_1, \cdots,  \phi^{\mathrm{v}}_{N_{\mathrm{v}}-1})$ and $\boldsymbol{\phi}_\mathrm{h}=(\phi^{\mathrm{h}}_0, \phi^{\mathrm{h}}_1, \cdots, \phi^{\mathrm{h}}_{N_\mathrm{h}-1})$ are assigned along vertical and horizontal lines, respectively.
The control value of each RIS element is represented as the product of the corresponding vertical and horizontal variables. The phase-shift vector of the entire RIS, $\boldsymbol{\phi}$, can be rewritten as
\begin{align}
\boldsymbol{\phi} = \mathrm{vec}(\boldsymbol{\phi}_\mathrm{h}^\mathrm{T}\boldsymbol{\phi}_\mathrm{v}),\label{eq:line_phase}
\end{align}
where $\mathrm{vec(\cdot)}$ denotes the vectorization of a matrix by stacking its columns.
This structure significantly reduces the number of control variables, simplifying the circuit and lowering implementation cost while maintaining sufficient beamforming flexibility.
In a conventional RIS system, all the $N_{\mathrm{v}}N_{\mathrm{h}}$ elements are optimized and controlled independently. Therefore, a dedicated control unit is required for each element. As the number of elements increases, the overall hardware cost and complexity grow substantially, making large-scale deployment impractical.
The optimization problem \eqref{eq:objective_ising} can be reformulated as
\begin{align}
    \begin{split}
        \max_{\bs\phi_\mathrm{h},\bs\phi_\mathrm{v}}& \quad (\vec(\bs{\phi}_{\mathrm{h}}^\mathrm{T}\bs\phi_{\mathrm{v}}))^\mathrm{H}\mathbf{J}(\vec(\bs{\phi}_{\mathrm{h}}^\mathrm{T}\bs\phi_{\mathrm{v}})),\\
        \textrm{s.t.}& \quad \phi^{\mathrm{v}}_i,\phi^\mathrm{h}_j \in\{-1,+1\} \quad \forall i,j.\label{eq:objective_line}
    \end{split}
\end{align}
This leads to a fourth-order objective function that cannot be directly solved using QA or CIM. Therefore, the objective function must be converted into a second-order one. The standard quadratization method in \cite{mandal2020compressed} requires manually chosen penalty parameters, and their choice strongly affects the resulting optimization performance. In addition, quadratization introduces computational overhead that becomes more severe for large problems. To overcome these limitations of the standard quadratization method, we propose a two-step optimization method. The proposed method reduces the problem size to approximately one-half of that in the standard method. Moreover, because it does not require manually tuned penalty parameters, it improves feasibility and scalability for larger problems.

\paragraph{First-step}
Here, we first focus on the two-level discretization of the phase shifts. Since the fourth-order problem in \eqref{eq:objective_line} cannot be directly handled by the CIM, we first ignore the line-control structure in \eqref{eq:line_phase} and optimize the full-RIS-element phases by solving
\begin{align}
\begin{split}
\max_{\bs\phi} \ & \boldsymbol{\phi}^{\mathrm H}\mathbf{J}\boldsymbol{\phi},\\
\text{s.t. } & \phi_i \in\{-1,+1\}\ \forall i.
\end{split}
\end{align}
We denote the obtained solution as a temporary solution $\boldsymbol{\phi}^*$.

\paragraph{Second-step}
Based on $\boldsymbol{\phi}^*$, we then fit a line-control pattern to $\boldsymbol{\phi}^*$ by solving
\begin{align}
\begin{split}
\min_{\bs\phi_\mathrm{h},\bs\phi_\mathrm{v}} \ &
\left\|
\boldsymbol{\phi}^* - \mathrm{vec}(\boldsymbol{\phi}_{\mathrm h}^{\mathrm T}\boldsymbol{\phi}_{\mathrm v})
\right\|^2,\\
\text{s.t. } &
\phi^{\mathrm v}_i,\phi^{\mathrm h}_j \in\{-1,+1\}\ \forall i,j.
\end{split}
\end{align}
Since $\boldsymbol{\phi}^*$ and $\mathrm{vec}(\boldsymbol{\phi}_{\mathrm h}^{\mathrm T}\boldsymbol{\phi}_{\mathrm v})$ belong to $\{-1,+1\}^{N}$, their squared norms are $N$.
Therefore, expanding the squared norm yields
\begin{align}
\left\|
\boldsymbol{\phi}^* - \mathrm{vec}(\boldsymbol{\phi}_{\mathrm h}^{\mathrm T}\boldsymbol{\phi}_{\mathrm v})
\right\|^2
= 2N - \Re\{2\,{\boldsymbol{\phi}^*}^{\mathrm H}\mathrm{vec}(\boldsymbol{\phi}_{\mathrm h}^{\mathrm T}\boldsymbol{\phi}_{\mathrm v})\},
\end{align}
and the problem is equivalently rewritten as
\begin{align}
\max_{\bs\phi_\mathrm{h},\bs\phi_\mathrm{v}} \ \Re\{{\boldsymbol{\phi}^*}^{\mathrm H}\mathrm{vec}(\boldsymbol{\phi}_{\mathrm h}^{\mathrm T}\boldsymbol{\phi}_{\mathrm v})\}.
\label{eq:second_step_corr}
\end{align}
Finally, using $\bigl[\mathrm{vec}(\boldsymbol{\phi}_{\mathrm h}^{\mathrm T}\boldsymbol{\phi}_{\mathrm v})\bigr]_{iN_{\mathrm h}+j}
=\phi^{\mathrm v}_i\phi^{\mathrm h}_j$, we obtain the quadratic objective
\begin{align}
\max_{\bs\phi_\mathrm{h},\bs\phi_\mathrm{v}}
\sum_{i=0}^{N_{\mathrm v}-1}\sum_{j=0}^{N_{\mathrm h}-1}
\phi^{*}_{iN_{\mathrm h}+j}\,\phi^{\mathrm v}_i\,\phi^{\mathrm h}_j.
\label{eq:two-step_1bit}
\end{align}
Thus, the resulting objective is quadratic and can be handled by physical machines such as QA and CIM.
The proposed method eliminates the need for manually defined parameters, enabling fast and stable optimization.

Following the analysis for the two-level case, the four-level case can be handled in the same manner by redefining the variables as
$\bs\phi_\mathrm{v}=1/\sqrt{2}(\bs\sigma_\mathrm{v,Re}+j\bs\sigma_\mathrm{v,Im})$,
$\bs\phi_\mathrm{h}=1/\sqrt{2}(\bs\sigma_\mathrm{h,Re}+j\bs\sigma_\mathrm{h,Im})$.
The number of optimization variables is doubled. This substitution makes the formulation equivalent to that of the two-level case, allowing the same technique to be applied directly.

\section{Performance Comparisons}
\label{sec:comp}

In this section, we compare the proposed two-step optimization method with the standard quadratization method. We also evaluate the received power obtained with a RIS optimized by an actual CIM system \cite{honjo2021100000spin} whose experimental settings are the same as those in \cite{takesue2025finding}. The Ising matrix was provided to the CIM after 8-bit quantization, where all $J_{ij}$ were uniformly scaled, rounded, and clipped to the signed range $[-127,127]$.

\subsection{Performance Evaluation of Each Quadratization Method}

We evaluated the proposed two-step optimization method for line-controlled phase shifts with $L=2$ and compared its performance with that of the optimal exhaustive search and the standard quadratization method. In the standard quadratization method, the penalty weights must be appropriately tuned, which introduces additional optimization overhead. For the standard method, we employed the tree-structured Parzen estimator (TPE) algorithm implemented in Optuna \cite{akiba2019optuna}, a hyperparameter optimization framework, to optimize these parameters over 200 trials for each setting.

\begin{figure}[tb]
\centering
\includegraphics[width=0.95\linewidth]{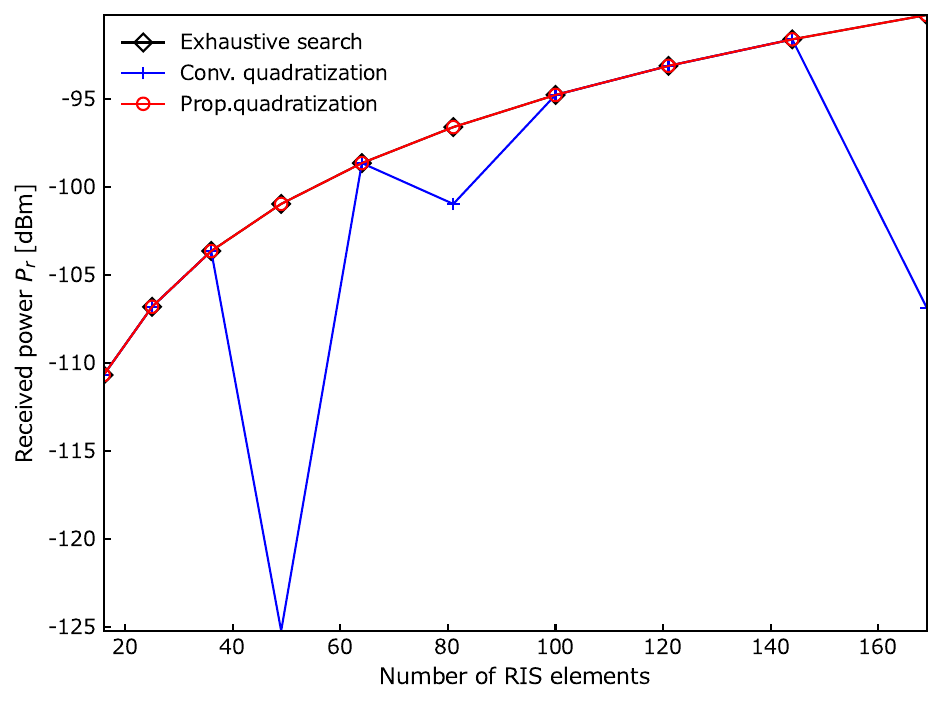}
\caption{Comparison of quadratization methods with exhaustive search, where the number of line-control variables is increased from 8 to 26.}
\label{fig:reduction}
\end{figure}

Figure~\ref{fig:reduction} shows the received power versus the number of variables. As shown, the proposed method achieved performance equivalent to the theoretical optimum. Furthermore, the proposed method reduced the number of variables after quadratization. Assuming a square RIS with $N_{\mathrm h}=N_{\mathrm v}=\sqrt{N}$, the standard method required approximately $2N$ variables on average, whereas the proposed two-step optimization reduced the number to $N$. For example, for a RIS with $74 \cdot 74 = 5,476$ elements, the required number of discrete variables was reduced from $11,100$ to $5,476$.

\begin{table}[tb]
\caption{Simulation Parameters}
\centering
\label{fig:sim_pram}
\begin{tabular}{|c|c|} \hline
Parameter & Value \\ \hline
Frequency $f$ & $28.0$ GHz \\ \hline 
Number of BS antennas & $64~(8\times8,$ half-wavelength spacing) \\ \hline 
Transmission power & $1$ W \\ \hline
Number of receive antennas & $1$ \\ \hline 
Number of RIS elements & from $16$ to $5,476$ \\ \hline 
\end{tabular}
\end{table}

\subsection{Performance Evaluation of Phase Optimization}

\begin{figure}[tb]
\centering
\includegraphics[width=1\linewidth]{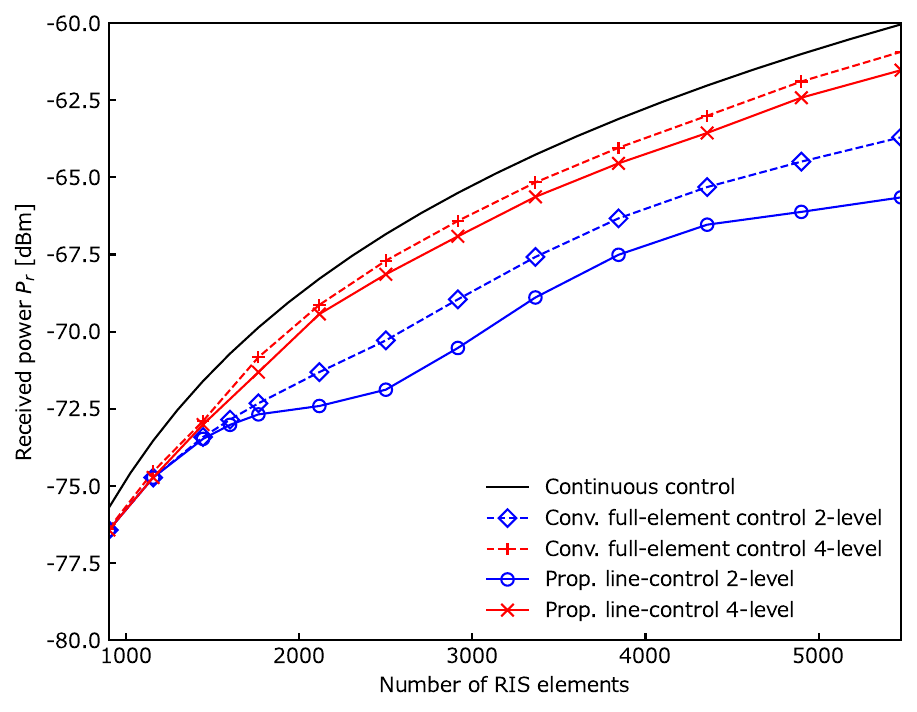}
\caption{Comparison of received power under different phase-control methods for varying numbers of RIS elements.}
\label{fig:risan_kekka}
\end{figure}

\begin{figure}[tb]
\centering
\includegraphics[width=1\linewidth]{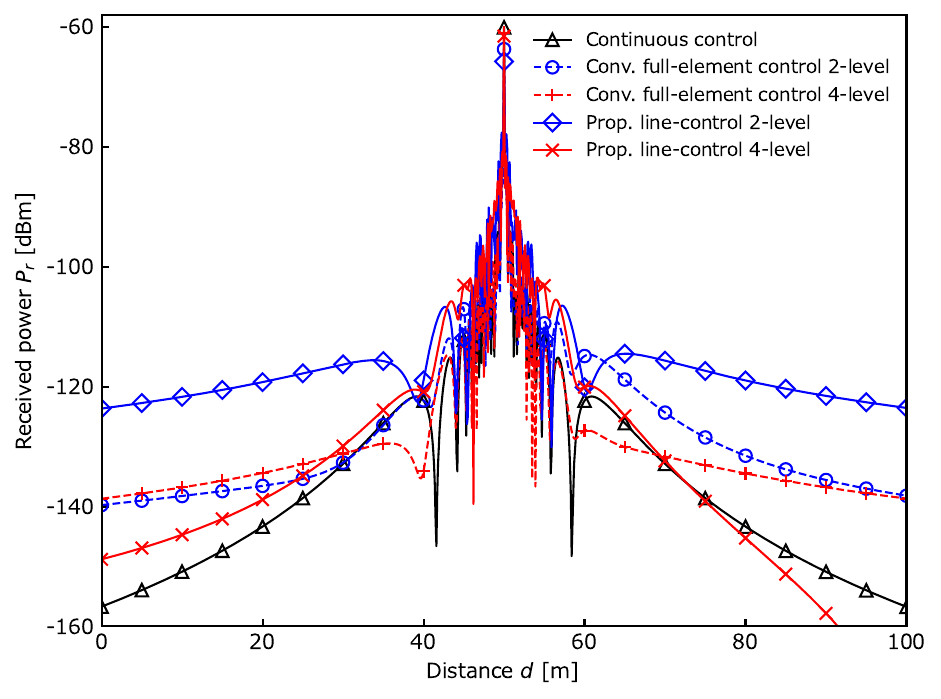}
\caption{Comparison of received power for 5,476 RIS elements at different distances.}
\label{fig:power_5476}
\end{figure}

We evaluated discrete phase control with $L=2$ and $L=4$. First, we compared the proposed line-controlled phase shifts with full-element-controlled phase shifts while varying the number of RIS elements. As a reference, we also included the theoretical upper bound given by continuous phase control.

As shown in Fig.~\ref{fig:risan_kekka}, all discrete phase-control methods achieved comparable performance when the number of RIS elements was small, whereas clear differences emerged as the number increased. At the design distance $d=50$ m, the proposed line-controlled phase shifts exhibited a performance gap of about 2 dB from the theoretical upper bound under continuous phase control for $L=2$, while this gap was reduced to less than 1 dB for $L=4$. Furthermore, Fig.~\ref{fig:power_5476} shows that increasing $L$ from 2 to 4 not only improved the received power at the design point but also effectively suppressed power leakage to undesired locations, which is beneficial for reducing inter-user interference and improving the sum rate in multi-user environments.

\section{Conclusions}
\label{sec:conc}
In this letter, we have proposed a two-dimensional line-control method for RIS phase shifts and a two-step method for mapping higher-order Ising problems onto quadratic problems that can be handled by physical machines such as the CIM. For efficient RIS control, we reduced the number of control elements from $N = N_\mathrm{v}N_\mathrm{h}$ to $N_{\mathrm{v}}+N_{\mathrm{h}}$. This reduction simplifies hardware implementation and facilitates the widespread adoption of large-scale RIS systems. Slight performance degradation was found to be inevitable because of the limited number of phase states imposed by hardware constraints. Nevertheless, this issue can be mitigated by increasing the discretization level from two to four levels. For fourth-order Ising problems, we also developed a penalty-free quadratization method that reduces the number of variables by approximately half compared with the conventional method. 
We experimentally validated the practical applicability of the proposed methods for RIS optimization with a large number of elements.

\footnotesize{
	\bibliographystyle{IEEEtranURLandMonthDeactivated}
    \bibliography{re_main,addition}
}

\end{document}